\documentclass[aps,prl,twocolumn,superscriptaddress]{revtex4}

\usepackage[dvips]{graphicx}
\begin{document}


\title{Coupling Of The B$_{1g}$ Phonon To The Anti-Nodal Electronic States of Bi$_{2}$Sr$_{2}$Ca$_{0.92}$Y$_{0.08}$Cu$_{2}$O$_{8+\delta}$}


\author{T. Cuk}
\author{F. Baumberger}
\author{D.H. Lu}
\author{N. Ingle}
\author{X.J. Zhou}
\affiliation{Departments of Physics, Applied Physics and Stanford
Synchrotron Radiation Laboratory, Stanford University, Stanford,
CA 94305}
\author{H. Eisaki}
\affiliation{Nanoelectronics Research Institute, National
Institute of Advanced Industrial Science and Technology, 1-1-1
Central 2, Umezono, Tsukuba, Ibaraki, 305-8568, Japan}
\affiliation{Departments of Physics, Applied Physics and Stanford
Synchrotron Radiation Laboratory, Stanford University, Stanford,
CA 94305}
\author{N. Kaneko}
\affiliation{Departments of Physics, Applied Physics and Stanford
Synchrotron Radiation Laboratory, Stanford University, Stanford,
CA 94305}
\author{Z. Hussain}
\affiliation{Advanced Light Source, Lawrence Berkeley National
Laboratory, Berkeley, California 94720, USA}
\author{T.P. Devereaux}
\affiliation{Dept. of Physics University of Waterloo Waterloo, ON
CANADA, N2L 3G1}
\author{N. Nagaosa}
\affiliation{CREST, Department of Applied Physics, University of
Tokyo, Bunkyo-ku, Tokyo 113, Japan}
\author{Z.-X. Shen}
\affiliation{Departments of Physics, Applied Physics and Stanford
Synchrotron Radiation Laboratory, Stanford University, Stanford,
CA 94305}
\affiliation{Physics Institute, University of Zurich,
Zurich, CH-8057, Switerland}


\date{October 5, 2003}

\begin{abstract}
Angle-resolved photoemission spectroscopy (ARPES) on optimally
doped Bi$_{2}$Sr$_{2}$Ca$_{0.92}$Y$_{0.08}$Cu$_{2}$O$_{8+\delta}$
uncovers a coupling of the electronic bands to a 40 meV mode in an
extended k-space region away from the nodal direction, leading to
a new interpretation of the strong renormalization of the
electronic structure seen in Bi2212. Phenomenological agreements
with neutron and Raman experiments suggest that this mode is the
B$_{1g}$ oxygen bond-buckling phonon. A theoretical calculation
based on this assignment reproduces the electronic renormalization
seen in the data.
\end{abstract}


\maketitle

The discovery of bosonic renormalization effects in cuprate
superconductors in the form of a dispersion ``kink" near 50-70 meV
for the nodal state has attracted considerable interest
\cite{kink1, kink2, kink3, kink1SR, kink2SR}. While a consensus
has been reached that the renormalization is due to electronic
coupling to a bosonic mode, disagreement remains as to whether the
assignment to a phononic or an electronic mode is more
appropriate.  More recently, another ``kink" phenomenon has been
reported for the antinodal electronic state near ($\pi$,0)
\cite{kink1SR, Gromko, Kim, Sato}. In contrast to the nodal
renormalization, which shows little change across T$_c$, the
antinodal renormalization has been observed so far only below
T$_{c}$. The strong temperature dependence and the dominance of
the coupling strength near ($\pi$,0) have been taken as evidence
to identify the bosonic mode with the 41meV spin resonance
\cite{kink1SR, Gromko, Kim, Sato}.

At first glance, the spin mode appears to provide a natural
interpretation for the antinodal kink. The spin mode turns on at
T$_c$ and has a well-defined momentum of ($\pi$,$\pi$) which
preferentially connects the anti-nodal states. However, serious
issues remain. The prominent kink in ARPES is observed in deeply
overdoped samples where no evidence of a spin resonance exists
\cite{Gromko,Kim}.  Further, there is a serious debate as to
whether the spin resonance has sufficient spectral weight to cause
the observed bosonic renormalization effect \cite{Kee, Abanov,
Sandvik}.

In this letter, we present extensive temperature, momentum, and
doping dependent ARPES data from
Bi$_{2}$Sr$_{2}$Ca$_{0.92}$Y$_{0.08}$Cu$_{2}$O$_{8+\delta}$ that
demonstrate the persistence of a dispersion break near 40meV in
the normal state close to ($\pi$,0), invalidating the spin
resonance interpretation. In the superconducting state, the energy
scale of this mode shifts to $\sim$ 65-70meV and the signatures of
coupling increase considerably.  We show that the $\sim$ 40meV
B$_{1g}$ oxygen ``bond-buckling" phonon has the correct energy and
coupling anisotropy from its d$_{x^{2}-y^{2}}$ symmetry
\cite{Reznik, TDevereaux2} that, in conjunction with the
underlying band-structure anisotropy, naturally explains the
momentum dependence of the ($\pi$,0) kink. We attribute the
temperature dependence to the density of states enhancement due to
the superconducting gap opening and to the thermal broadening of
the phonon self energy in the normal state. Calculations using an
electron-phonon coupling vertex of B$_{1g}$ symmetry reproduce the
experimental data.

ARPES experiments have been preformed at beamline V-4 of the
Stanford Synchrotron Radiation Laboratory with a photon energy of
22.7eV. The energy and angular resolutions are 14 meV and
$\pm$0.15 degree, respectively. Optimally doped
Bi$_{2}$Sr$_{2}$Ca$_{0.92}$Y$_{0.08}$Cu$_{2}$O$_{8+\delta}$(T$_{c}$
= 94K) samples were cleaved in situ with a base pressure of 4 x
10$^{-11}$ torr. Key data have been reproduced in two other
experimental chambers.

In Fig.\ 1, data in the anti-nodal region reveal a dramatic change
in the effective coupling through T$_{c}$. In the superconducting
state, classical Engelsberg-Schrieffer signatures of electronic
coupling to a bosonic mode are seen in the Energy Distribution
Curves (EDCs) (Fig. 1b1) and the image plot (Fig. 1b2): 1) a break
up into two branches---a peak that decays as it asymptotically
approaches a characteristic energy at 70meV, and a hump that
traces out a roughly parabolic band.  The peak that asymptotically
approaches the mode energy derives from strong mixing of the
electronic band with the mode, while the hump traces out the bare
band dispersion away from the mode energy (the peaks are indicated
by ``I" and the humps by ``II" in Fig. 1b1; the peaks correspond
to the trail of intensity above the indicated point in Fig. 1b2);
2) a significant broadening of the spectra beyond 70meV due to the
onset of the bosonic mode self energy\cite{Engels}.

\begin{figure}[htb]
\includegraphics[width=0.5\textwidth]{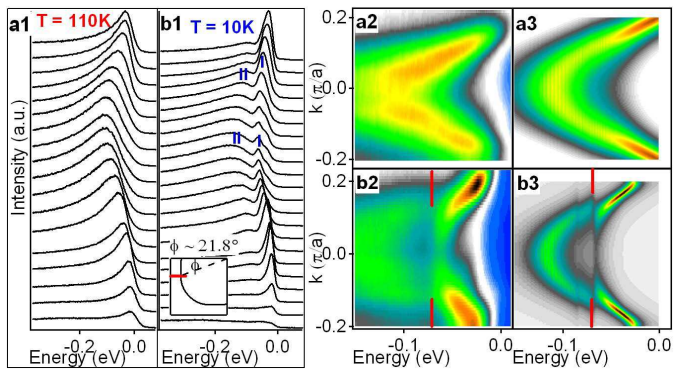}
\caption{\label{Figure1} a1) and b1) EDCs taken at 110K and 15K,
respectively, at the k-space location indicated in the inset of
b1). The corresponding image plots are shown in a2) and b2). Image
plots of the calculations are shown in a3) and b3) at 110K and 15K
respectively. Color scales are independent for each image plot.}
\end{figure}

In the normal state (Fig. 1a1, 1a2) little of these effects can be
seen without further analysis. In Fig. 2, we have extracted
dispersions for three k-space cuts in a momentum space region
between the nodal direction and the Van Hove Singularity (VHS) at
($\pi$,0). Since we are only concerned with the energy scale, we
fit the Energy Distribution Curves (EDCs) phenomenologically. The
usual method to extract dispersions---fitting Momentum
Distribution Curves with Lorentzians---is not appropriate since
the assumed linear approximation of the bare band fails towards
($\pi$,0) where the band bottom is close to E$_{F}$\cite{LaShell}.
EDC-derived dispersions of three independent data sets shown in
Fig. 2 (a1,b1,c) consistently reveal a $\sim$40meV energy scale
that has eluded detection before \cite{kink1SR,Gromko, Sato, Kim}.

One may be concerned about a distortion to the dispersion due to
bilayer splitting. However, the 40 meV mode has been clearly
observed in two other cases that are independent of bilayer
effects---in deeply over-doped samples below T$_{c}$ where the gap
is small, for which the bilayer splitting is well-resolved
\cite{Gromko}, and in normal state optics data near optimal doping
for which spectra are averaged over the Brillouin Zone, thereby
washing out any sharp structure due to the dispersion of bilayer
splitting \cite{Tu}. Further, while the bilayer splitting is
different under the three circumstances shown in Fig. 2 (a1, b1,
c), the kink energy remains the same. Finally, while certain
peak-dip-hump features have been attributed to bilayer splitting
when well-resolved, ``kink" features in the dispersions cannot be
accounted for in this way (two nearly parallel bands alone do not
reproduce the effect).

In Fig. 2, we also show the temperature dependence of this 40meV
kink through a comparison of the dispersions. In the
superconducting state, the EDC peak position asymptotically
approaches the characteristic energy defined by the bosonic mode
(Fig. 2a2, Fig. 2b2). Because the effect is much stronger below
T$_{c}$, it is clearly revealed in the MDC dispersion also; the
kink position indicates the characteristic energy and the kink
sharpness monotonically increases with the coupling strength. In
addition to the increase in effective coupling strength, the data
show that the kink energy shifts from $\sim$40meV to
$\sim$65-70meV. This is an important observation since the opening
of a superconducting gap is expected to shift the energy at which
the electronic states couple to the bosonic mode and has thus far
not been detected by ARPES. Here, we observe a kink shift of
$\sim$25-30meV, close to the maximum gap energy, $\triangle_{0}$.
We summarize the temperature dependence of the energy at which we
see a bosonic mode couple to the electronic states in the
anti-nodal region in Fig. 2(d): the kink energies are at
$\sim$40meV above T$_{c}$ near the anti-nodal region and increase
to $\sim$70meV below T$_{c}$. Because the band minimum is too
close to E$_{F}$, the normal state $\sim$40meV kink cannot easily
be seen below $\phi\sim20^{o}$.

\begin{figure}[htb]
\includegraphics[width=0.5\textwidth]{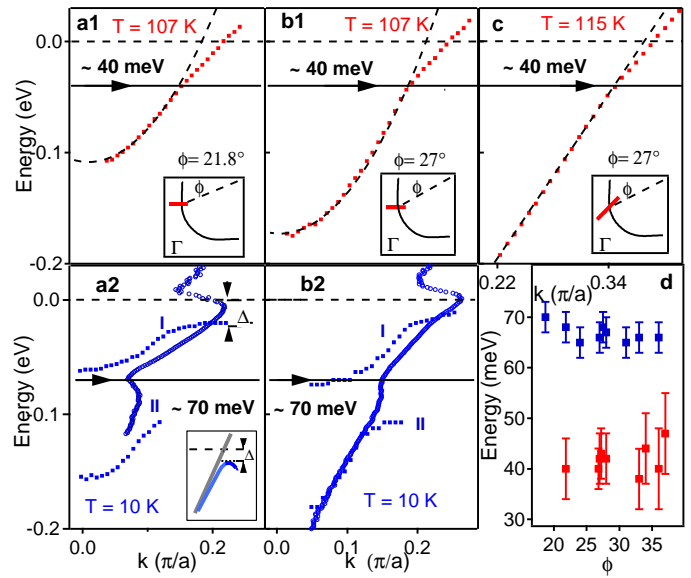}
\caption{\label{Figure2} EDC(a1, b1, c) derived dispersions in the
normal state (107K and 115K). $\phi$ and the cut-direction are
noted in the insets.  The red dots are the data; the fit to the
curve (dashes, black) below the 40meV line is a guide to the eye.
a2) and b2) are MDC derived dispersions at the same location and
direction as in a1) and b1), but in the superconducting state
(15K). In a2) and b2) we also plot the peak(I) and hump
positions(II) of the EDCs for comparison. The inset of a2) shows
the expected behavior of a Bogoliubov type gap opening. The s-like
shape below the gap energy is an artifact of how the MDC handles
the back-bend of the Bogoliubov quasiparticle. d) kink positions
as a function of $\phi$ in the anti-nodal region.}
\end{figure}

Now we turn to the momentum dependence shown in Fig. 3 that
reveals two features defining the anisotropy of the bosonic mode
coupling. First, the coupling is extended in the Brillouin zone
and has a similar energy scale throughout, near $\sim$ 65-70meV.
Second, the signatures of coupling increase significantly toward
$(\pi,0)$ or smaller $\phi$. The clear minimum in spectral weight
seen near 70meV in Fig. 3a1) and 3a2) indicates strong mixing of
the electronic states with the bosonic mode where the two bare
dispersions coincide in energy. In the same figure, we show how
these distinctive features change as a function of doping. In the
deeply overdoped sample (T $_{c}$ $\sim$ 65K or $\delta \sim$
22$\%$), Fig. 3b, the kink energy moves to 40meV since
$\Delta_{0}$ ($\sim$ 10-15meV) becomes much smaller. In the
underdoped sample (T $_{c}$ $\sim$ 85K), the kink energy remains
around 70meV since it has a similar gap as the optimally doped
sample ($\Delta_{0}$ $\sim$ 35-40meV). The signatures of coupling
remain strong throughout, although they do noticeably increase
from the overdoped to underdoped sample when comparing data taken
at the same $\phi$.

\begin{figure*}[htb]
\includegraphics[width= 1\textwidth]{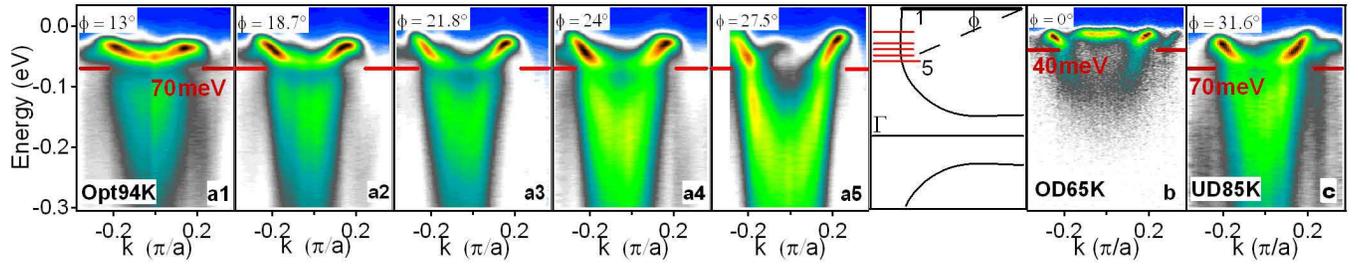}
\caption{\label{Figure3} The image plots in a1-a5)are cuts taken
parallel to (0,$\pi$)-($\pi$,$\pi$) at the locations indicated in
the zone at 15K for an optimally doped sample (94K). b) and c) are
spectra taken parallel to (0,$\pi$)-($\pi$,$\pi$) at the k-space
locations indicated for over-doped (65K) and under-doped samples
(85K) respectively.}
\end{figure*}

The above described experimental observations severely constrain
the applicability of the spin resonance mode: 1.) We observe the
40meV boson in the normal state at optimal doping while the 41meV
spin resonance mode exists only below T$_{c}$ \cite{Fong,Dai} 2.)
The kink is sharp in the super-conducting state of a deeply
over-doped Bi2212 sample (Fig. 3b, consistent with data of
\cite{Gromko}) where no spin resonance mode has been reported, or
is expected to exist since the spin fluctuation decreases strongly
with doping. 3.) The kink seen in under-doped Bi2212 (Fig. 3c) is
just as sharp as in the optimally doped case, while the neutron
resonance peak is much broader\cite{Fong, Dai}. 4) Since the
mode's spectral weight is only 2$\%$ \cite{Kee}, it may only cause
a strong enough kink effect by highly concentrating in k-space
\cite{Abanov}. Our data argues against this point as the bosonic
renormalization effects exist in an extended k-space range.

We now propose an alternative interpretation for the anti-nodal
renormalization effect as due to a coupling of the electronic
states with the near 40meV B$_{1g}$ phonon involving the
out-of-plane motion of the in-plane oxygen \cite{Reznik}. Neutron
scattering experiments show that this mode exhibits a clear
softening across T$_{c}$ only for q $<$ 0.5 $\pi$/a \cite{Reznik}.
As depicted by the red arrows in Fig. 4a), this is consistent with
a phonon coupling primarily states near the anti-node that can be
connected with small q. A more complete picture, which we reserve
for future discussion, would need to consider the 70 meV
half-breathing phonon shown to couple to the nodal states
\cite{kink2} and known to soften with doping by large q
\cite{Chaplot}.

The question now is whether the assignment to the B$_{1g}$ phonon
can explain the coupling anisotropy and temperature dependence. We
address the momentum dependence with a re-formulation of a
previous theory \cite{TDevereaux}. The B$_{1g}$ phonon
preferentially couples to $k$ points in the anti-nodal region of
the BZ \cite{TDevereaux2, Opel}. Fig. 4a) and Fig. 4b) show the
magnitude squared of the electron-phonon coupling vertex,
g$^{2}$($\textit{k}$,$\textit{k'}$), as a function of both the
initial momentum of the electron, $\textit{k}$, and the phonon
momentum, $\textit{q}$, connecting $\textit{k}$ to $\textit{k'}$ =
$\textit{k}$ + $\textit{q}$ along the Fermi surface. An electron
initially at the anti-node ($\textit{k$_{AN}$}$), Fig. 4a),
couples preferentially to other states in the anti-nodal region
and especially favors $\textit{q}\sim2k_{f}$ scattering. An
electron initially at the node ($\textit{k$_{N}$}$), Fig. 4b),
couples preferentially to states midway between the node and the
anti-node. A comparison of Fig. 4a) and 4b) further reveals that
g$^{2}$($\textit{k}$,$\textit{k'}$) decreases substantially for an
electron initially at the node. This momentum dependence in the
electron-phonon coupling suggests that the signatures of strongest
coupling should occur in cuts towards the anti-node, and parallel
to the $(\pi,0)$-$(\pi,\pi)$ line, as observed in the data. The
strong momentum dependence also explains why the coupling constant
inferred by ARPES measurements can be so different from the LDA
value ($\lambda \sim $ 0.3 for which the buckling mode dominates
\cite{Jepsen}) which represents an average of both $\textit{k}$
over the Brillouin zone and $\textit{q}$ over the momentum
transfer. ARPES measurements correspond to a $\lambda \sim $ 2.8
at the zone axes, but to an averaged $\lambda \sim $ 0.2 and an
even smaller value of the Raman $\lambda$ that represents
$\textit{q}$=0. Within the context of this calculation, which
relies on the framework of band theory, a quantitative comparison
with the electron-phonon vertex derived from fitting Raman
experiments \cite{Opel} cannot be made. In order to do so, one has
to consider vertex corrections which will affect the Raman and
ARPES-derived coupling differently \cite{Oliver}. We leave these
many body effects for future work.

\begin{figure}[bth]
\includegraphics[width=0.37\textwidth]{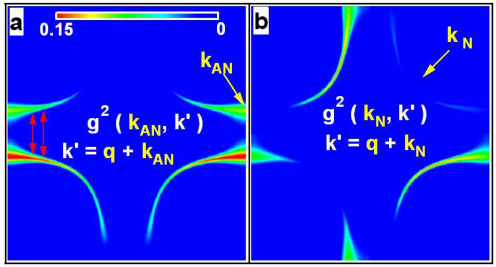}
\caption{\label{Figure4} a) and b) image plots of g$^{2}$(k,k')
for the B$_{1g}$ bond-buckling phonon on the Fermi Surface of
Bi2212. k' = k + q, where k is the initial momentum of the
electron and q is the momentum of the phonon. a) represents
g$^{2}$(k$_{AN}$,k') for an electron initially at the anti-node.
We draw red arrows indicating the dominant
q$\sim$2k$_{f}$$\sim$0.5$\pi/a$ scattering; b) represents
g$^{2}$(k$_{N}$,k') for an electron initially at the node}
\end{figure}

Finally, we turn to the temperature dependence. Similar to ARPES
data on Bi2212 reported here, tunneling data from a classical
superconductor such as Pb shows wiggles due to electron-phonon
coupling in the superconducting state, while they are hardly
detectable in the normal state \cite{Pbwiggles}. Motivated by the
case of Pb, and other strongly-coupled conventional
superconductors, we have solved the Eliashberg equations using a
one-step iteration procedure as described in Sandvik et al.
\cite{Sandvik}. The self energy includes an electron-phonon
coupling of the Bi2212 electronic structure described by a tight
binding model fit to the measured dispersion and a d-wave gap in
the superconducting state to the 40 meV B$_{1g}$ phonon with the
g$^{2}$(k,k') vertex shown in Fig. 4. Line cuts of the calculation
close to the anti-node are compared with image plots of the data
in Fig. 1. The calculations show that the dramatic increase of the
effective coupling in the superconducting state can be attributed
to the density of states enhancement due to the opening of the
superconducting gap. The substantially higher temperature ($\sim$
100K) in the normal state also serves to broaden the phonon
feature so that the dispersion exhibits, at most, a ``kink" effect
rather than a break up into two bands. The complete analysis of
this calculation will be presented in a subsequent paper
\cite{TDevereaux}.

In summary, based on extensive temperature, momentum, and doping
dependent data and a theoretical calculation, we propose a new
interpretation of the bosonic mode renormalization seen in ARPES
as electronic coupling to the B$_{1g}$ bond-buckling phonon. The
dominance of the renormalization near the anti-node indicates its
potential importance to the pairing mechanism, which is consistent
with some theory \cite{Muller, Scalapino2, Jepsen, Nazarenko} but
remains to be investigated.

\end{document}